\documentclass[preprint]{emulateapj}
\usepackage{natbib}
\usepackage{graphicx}

\begin{document} 

\title{Direct imaging of the water snow line at the time of planet formation using two ALMA continuum bands} 

\author{A. Banzatti\altaffilmark{1}, P. Pinilla\altaffilmark{2}, L. Ricci\altaffilmark{3}, K. M. Pontoppidan\altaffilmark{1}, T. Birnstiel\altaffilmark{3}, F. Ciesla\altaffilmark{4}} 
\altaffiltext{1}{Space Telescope Science Institute, 3700 San Martin Drive, Baltimore, MD 21218, USA} 
\altaffiltext{2}{Leiden Observatory, Leiden University, P.O. Box 9513, 2300RA Leiden, The Netherlands} 
\altaffiltext{3}{Harvard-Smithsonian Center for Astrophysics, 60 Garden Street, Cambridge, MA 02138, USA} 
\altaffiltext{4}{Department of the Geophysical Sciences, The University of Chicago, 5734 South Ellis Avenue, Chicago, IL 60637, USA} 

\email{banzatti@stsci.edu}

\begin{abstract} 
Molecular snow lines in protoplanetary disks have been studied theoretically for decades because of their importance in shaping planetary architectures and compositions. The water snow line lies in the planet formation region at $\lesssim10$\,AU, and so far its location has been estimated only indirectly from spatially-unresolved spectroscopy. This work presents a proof-of-concept method to directly image the water snow line in protoplanetary disks through its physical and chemical imprint in the local dust properties. We adopt a physical disk model that includes dust coagulation, fragmentation, drift, and a change in fragmentation velocities of a factor 10 between dry silicates and icy grains as found by laboratory work. We find that the presence of a water snow line leads to a sharp discontinuity in the radial profile of the dust emission spectral index $\alpha_{\rm{mm}}$, due to replenishment of small grains through fragmentation. We use the ALMA simulator to demonstrate that this effect can be observed in protoplanetary disks using spatially-resolved ALMA images in two continuum bands. We explore the model dependence on the disk viscosity and find that the spectral index reveals the water snow line for a wide range of conditions, with opposite trends when the emission is optically thin rather than thick. If the disk viscosity is low ($\alpha_{\rm{visc}} < 10^{-3}$) the snow line produces a ring-like structure with a minimum at $\alpha_{\rm{mm}}\sim2$ in the optically thick regime, possibly similar to what has been measured with ALMA in the innermost region of the HL Tau disk.
\end{abstract}


\section{INTRODUCTION} \label{sec:intr}
The structure of gas-rich protoplanetary disks is divided into icy and ice-free regions, depending on the local temperature of dust grains. The delimiting surfaces between these regions are known as ``condensation fronts" or ``snow lines". Each molecular species has its own condensation front, depending on its unique freeze-out temperature, but the water snow line is the dominant (in volatile mass) and innermost one. The evolving location of the water snow line affects the efficiency of planetesimal \citep{Blum08}, super-Earth \citep{Howe15} and giant planet core formation \citep{Kennedy08,morbi15}, the composition of giant planet atmospheres \citep{Oberg11,Madhusudhan14} as well as the delivery of water to the surfaces of terrestrial planets \citep[e.g.][]{Raymond07}. Numerous models have calculated its location in protoplanetary disks, often with conflicting predictions \citep[e.g.][]{Lecar06,Garaud07,Oka11,Martin12}. Directly observing where the water snow line is, and how it evolves during the time of planet formation, is therefore critical for validating and improving our understanding of the initial conditions of (exo)planetary systems.

Imaging a snow line in a disk, however, presents a number of observational challenges. Essentially, $i)$ the existence of an observable tracer for icy solids in the disk midplane, and $ii)$ the angular extent in the sky of the snow line region as seen from Earth. The former determines the spectral setting to use, the latter defines the angular resolution to achieve. Both these constraints cooperate constructively in the case of the CO snow line: gas emission tracers activated by CO ice, coupled to the low condensation temperature of CO pushing its snow line to radii $> 20\,$AU \footnote{An angular region of $\gtrsim0\farcs4$ if observed from 100 pc away}, recently provided resolved images by the Atacama Large Millimeter/Submillimeter Array (ALMA) in two protoplanetary disks \citep{qi15}.
The case of water, instead, is made more challenging by a higher condensation temperature of $\approx150\,$K, which puts its snow line at only a few AU in disks around solar-mass stars \citep[e.g.][]{Mulders15}. While the distribution of water vapor has been estimated in disk surfaces using infrared spectroscopy \citep{zhang13,sandra}, the location of the snow line in the disk midplane, where planetesimal formation occurs, has never been directly observed. 

Models of the water distribution in evolving protoplanetary disks suggest that its snow line may imprint signatures in the dust surface density \citep{CC06,Til10}. As mm/cm-size pebbles decouple from the gas and radially drift towards the pressure maximum \citep{weid77}, the composition and size distribution of dust grains are expected to change dramatically. Beyond the snow line, grains are largely composed of ice, and can efficiently coagulate into larger pebbles \citep{brauer08a,ros13}. Ice evaporation, instead, decreases the velocity threshold above which particles suffer destructive collisions (the ``fragmentation" velocity), leading to enhanced production of small grains \citep{Til10}. 

In this work, we model the evolution of dust grains in the midplane of a protoplanetary disk structure, to determine whether the water snow line can be detected by ALMA continuum imaging. We find that a snow line causes a radial discontinuity in the dust grain sizes and imprints a strong signal in the dust thermal emission (Section \ref{sec:modeling}). We demonstrate that this can be observed by ALMA as an inversion in the inward decreasing trend of the dust emission spectral index (Section \ref{sec:alma_sim}).

\section{Modeling the snow line imprint in the dust} \label{sec:modeling}
\begin{figure}
\centering
\includegraphics[width=0.45\textwidth]{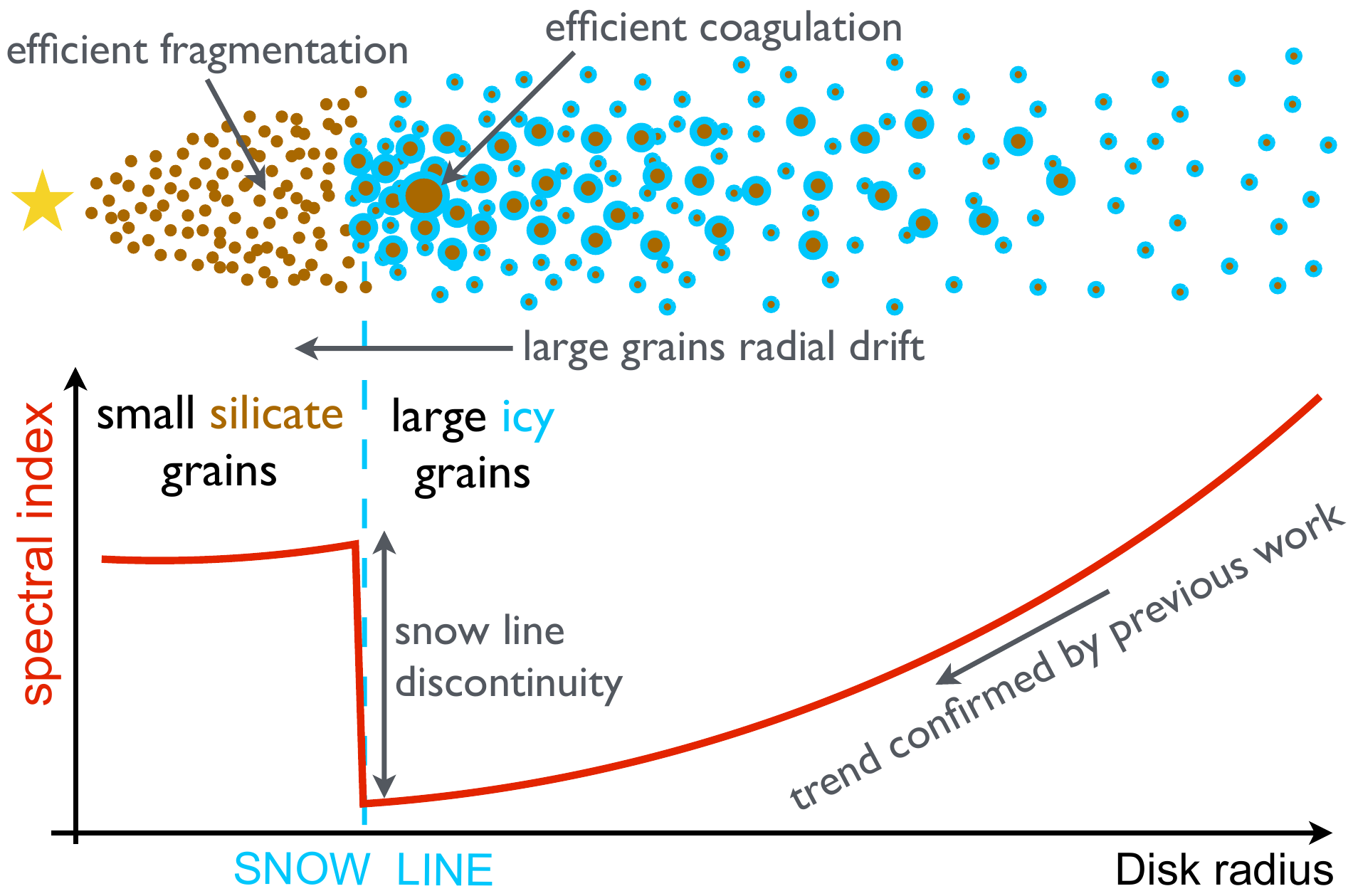} 
\caption{Illustrative sketch (not to scale) of the change in the dust emission spectral index $\alpha_{\rm{mm}}$ as linked to the change in dust properties across the water snow line, in the case of optically thin disk emission at millimeter wavelengths (see Section \ref{sec: model_results}).}
\label{fig:cartoon}
\end{figure}

The dust thermal emission from disks at millimeter wavelengths follows a power-law $F_{\nu} \propto \nu^{\alpha_{\rm mm}}$ \citep{beck90}. In the case of optically thin emission, the dust emission spectral index $\alpha_{\rm{mm}}$ is linked to the dust opacity $k_{\nu} \propto \nu^{\beta}$ as $\alpha_{\rm mm} = 2 + \beta$. The index $\beta$ is linked to the dust grain properties, including their size and their composition, through their opacity, where grains with diameters $a \gg \lambda/\pi$ have mass opacities that decrease with particle size \citep[e.g.][]{isella10,testi14,andr15}. Conversely, in case of optically thick emission, $F_{\nu} \propto \nu^{2}$ in the Rayleigh-Jeans regime. That is, $\alpha_{\rm mm} = 2$ while $\beta$ remains unconstrained because the emission is not sensitive to the dust grain properties. Millimeter interferometric images of ``classical" disks have provided evidence for a decrease of $\alpha_{\rm mm}$ towards smaller disk radii \citep{isella10,banz,guill,perez}. Measurements of $\alpha_{\rm mm}<3$ inside of 50 AU (limited by the available spatial resolution of $\gtrsim 20$ AU) for optically thin disks indicated that the emission is dominated by mm-cm-size grains ($\beta<1$). At larger disk radii, smaller grains dominate the emission ($\alpha_{\rm mm}>3.5$). These observations have been interpreted as radial segregation of solids, as produced by radial drift due to gas drag \citep{weid77,Til14}. In this work we demonstrate that, in addition to this trend, a sharp and strong discontinuity in the spectral index profile is expected at the snow line, where the innermost part of the disk is replenished with small grains through fragmentation (Figure \ref{fig:cartoon}).

To quantify the change of $\alpha_{\rm{mm}}$ across the snow line, we model the evolution of dust in a protoplanetary disk including the effects of radial drift, coagulation, and fragmentation of dust grains, as implemented in series of papers e.g. by \citet{pin12,pin15}, based on one-dimensional dust evolution models by \citet{brauer08b} and \citet{Til10}.
We assume that the dust population is initially comprised of $\mu$m-size particles, well-mixed with the gas in a disk with a radially constant gas-to-dust ratio of 100. During evolution, the particles stick together and aggregate depending on their relative velocities, including Brownian motion, radial drift, vertical settling, and turbulent mixing. The radial motion of the particles is size-dependent, and is computed taking into account drag, drift and turbulent diffusion \citep{brauer08b}. Particles stick when their relative velocities are lower than the fragmentation velocity ($v_{\rm{frag}}$), which depends on their composition, otherwise they fragment. 

In the model, the maximum dust grain size ($a_{\rm{frag}}$) achievable at a given disk radius $r$ is given by \citep{Til10,Til12}

\begin{equation} \label{eqn: a_max}
a_{\rm{frag}} (r) \propto \frac{\Sigma_{g} (r)}{\rho_s \alpha_{\rm{visc}}} \, \frac{v^{2}_{\rm{frag}}(r)}{c_s^2(r)} \, ,
\end{equation} 
where $\Sigma_{g} (r)$ is the gas surface density, $\rho_s$ is the volume density of a single grain particle (typically of the order of $1~\rm{g~cm}^{-3}$), $c_s$ is the sound speed, and $\alpha_{\rm{visc}}$ is a dimensionless parameter to quantify the disk viscosity \citep{shaksun73}. Stronger dust diffusion leads to higher relative turbulent velocities ($v_{\rm{turb}}\propto \sqrt{\alpha_{\rm{visc}}} c_s$). Therefore higher values of $\alpha_{\rm{visc}}$ (or $c_s$) decrease the maximum grain size. On the other hand, the drift efficiency depends on the grain size \citep{brauer08b}. Under fragmentation conditions, the maximum grain size depends on $\alpha_{\rm{visc}}$, and there is an indirect relation between radial drift and $\alpha_{\rm{visc}}$, where the lower the $\alpha_{\rm{visc}}$ the more efficient the drift.

\begin{figure*}
\centering
\includegraphics[scale=0.3]{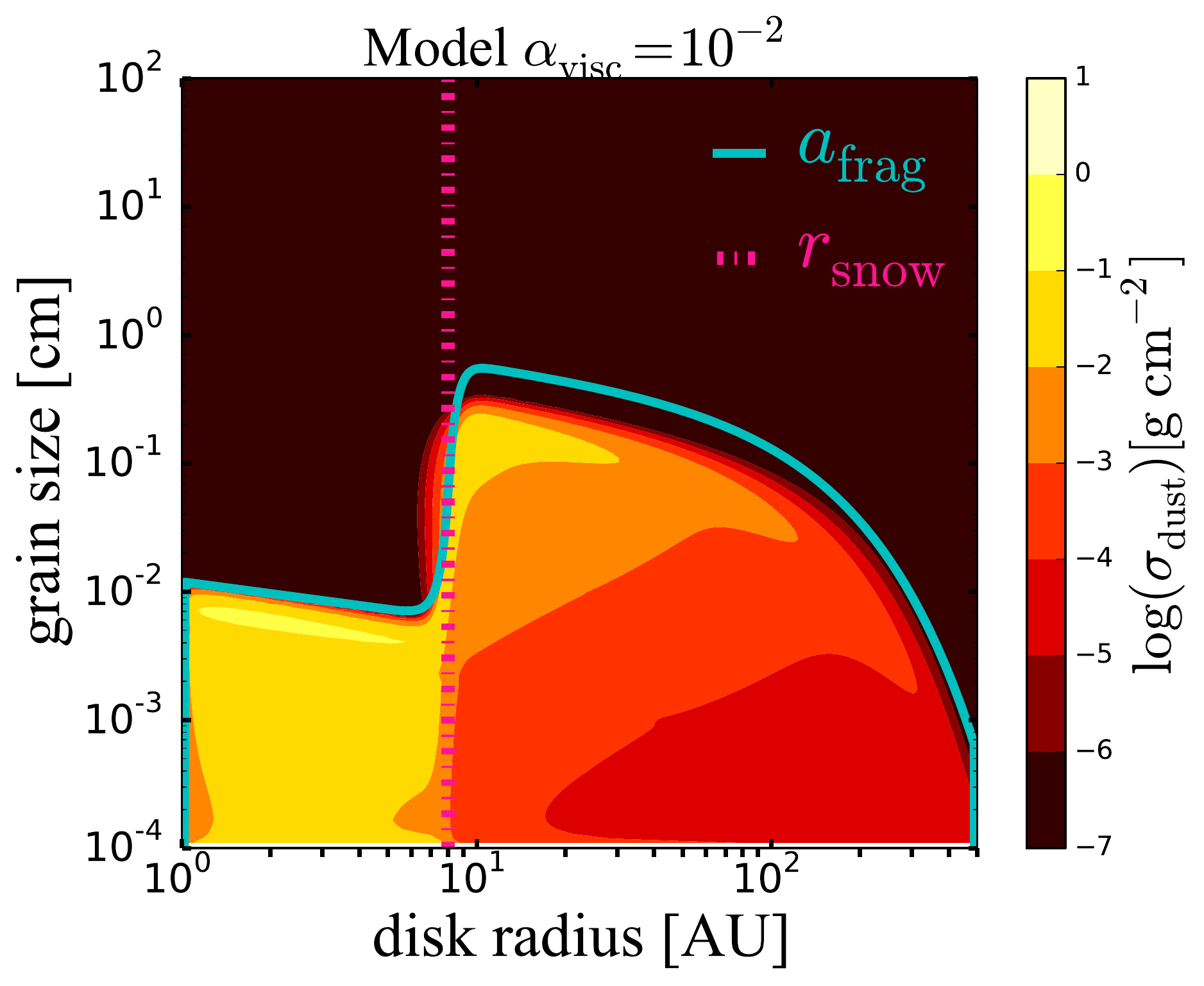}
\includegraphics[scale=0.3]{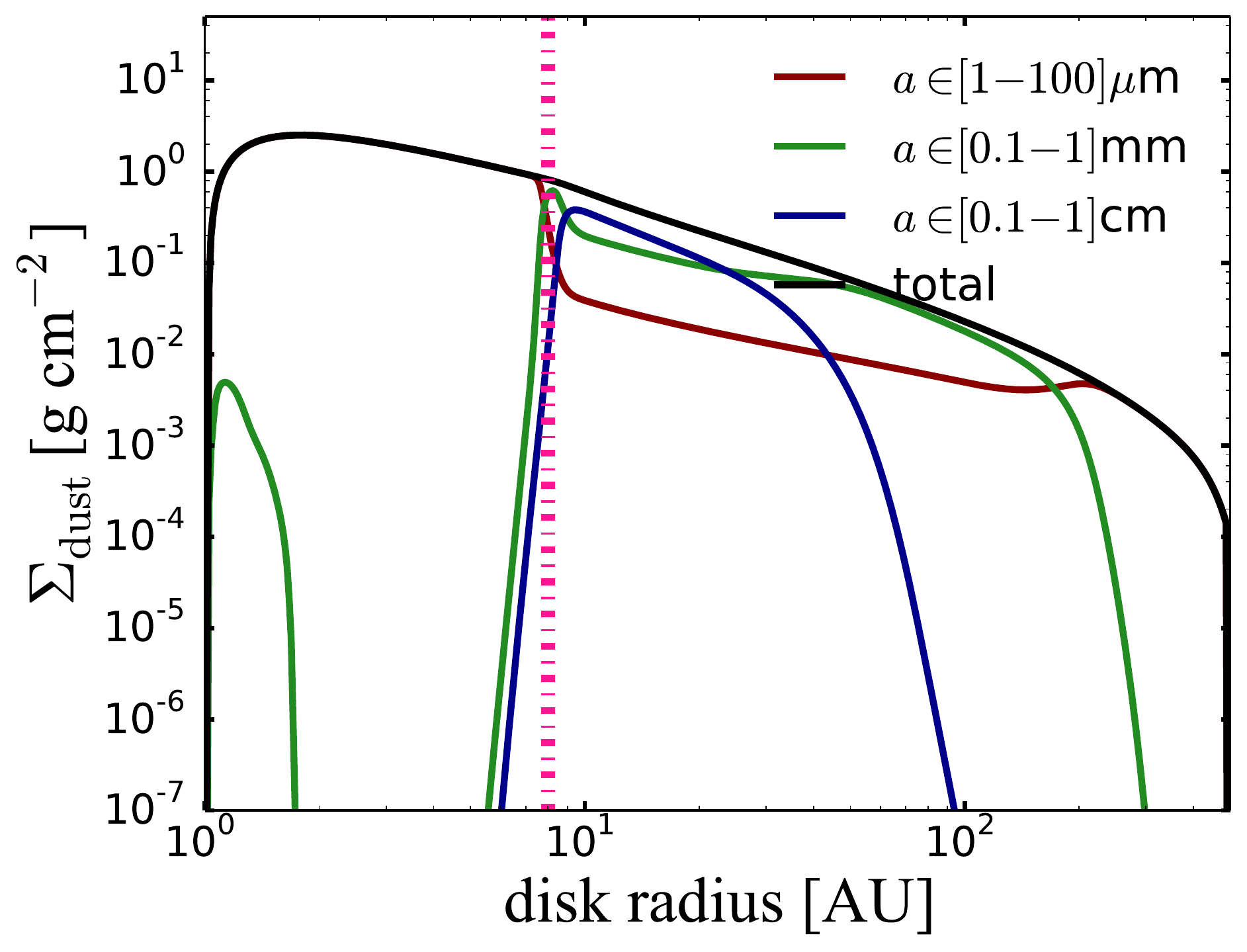}
\includegraphics[scale=0.3]{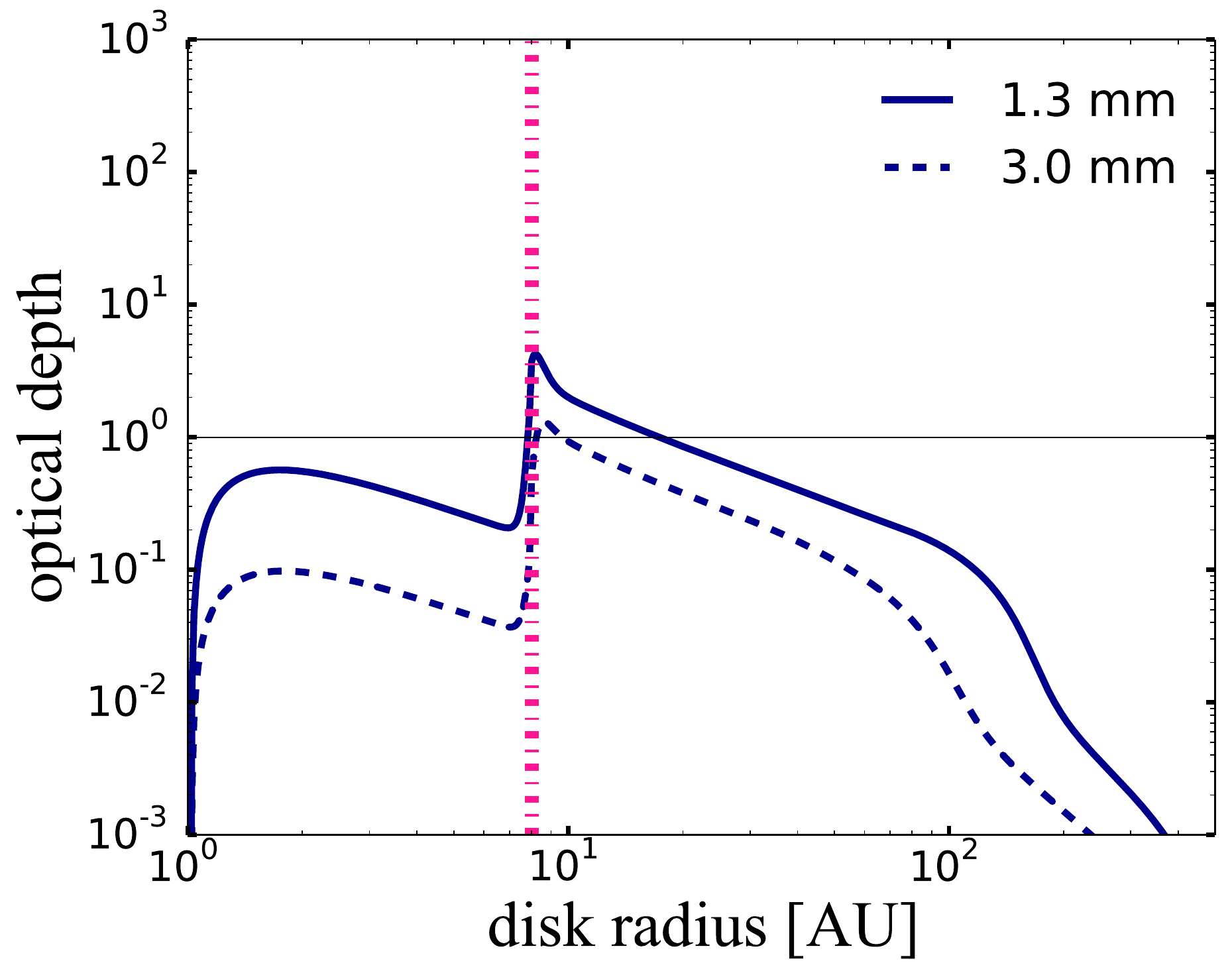}
\includegraphics[scale=0.3]{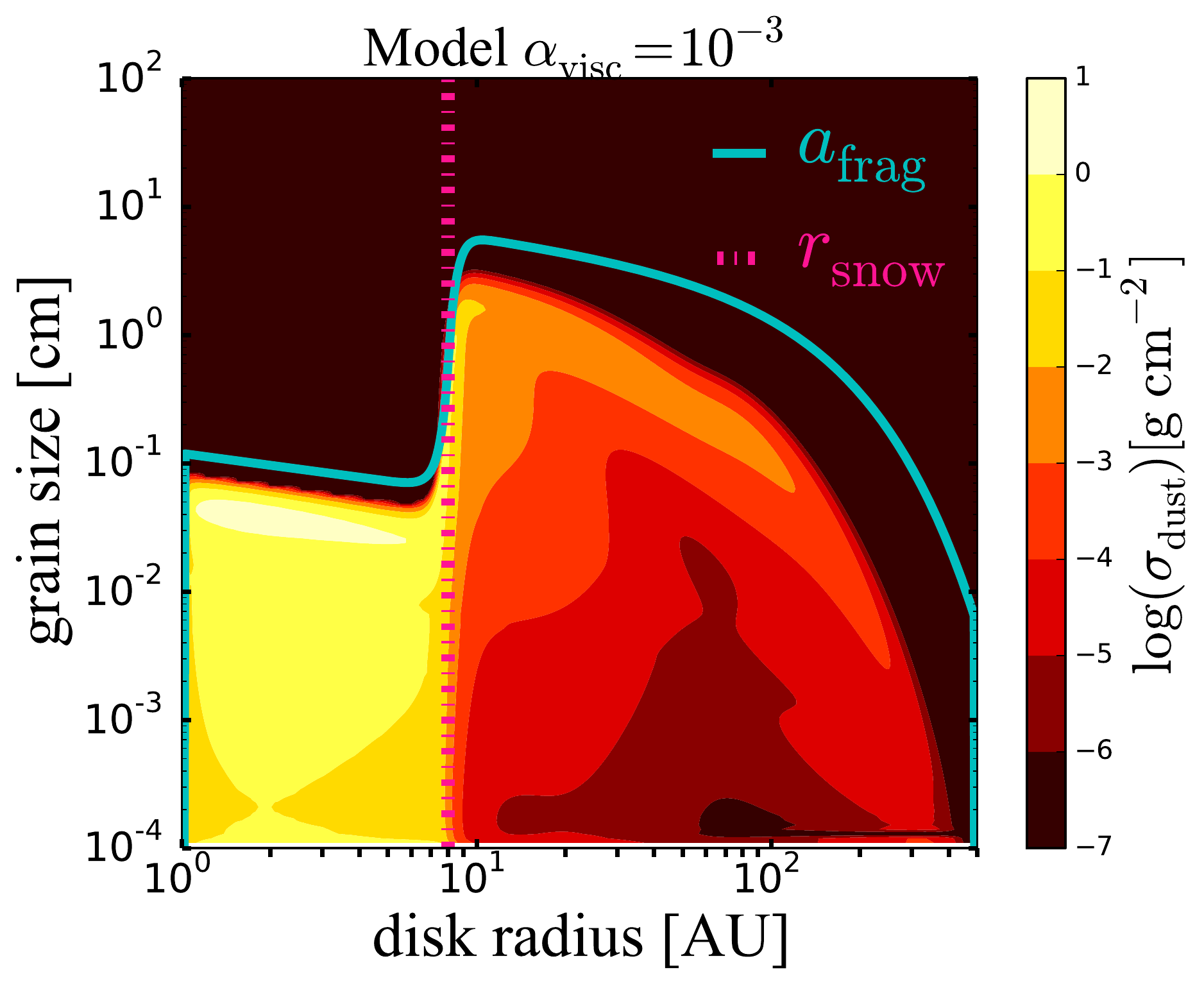}
\includegraphics[scale=0.3]{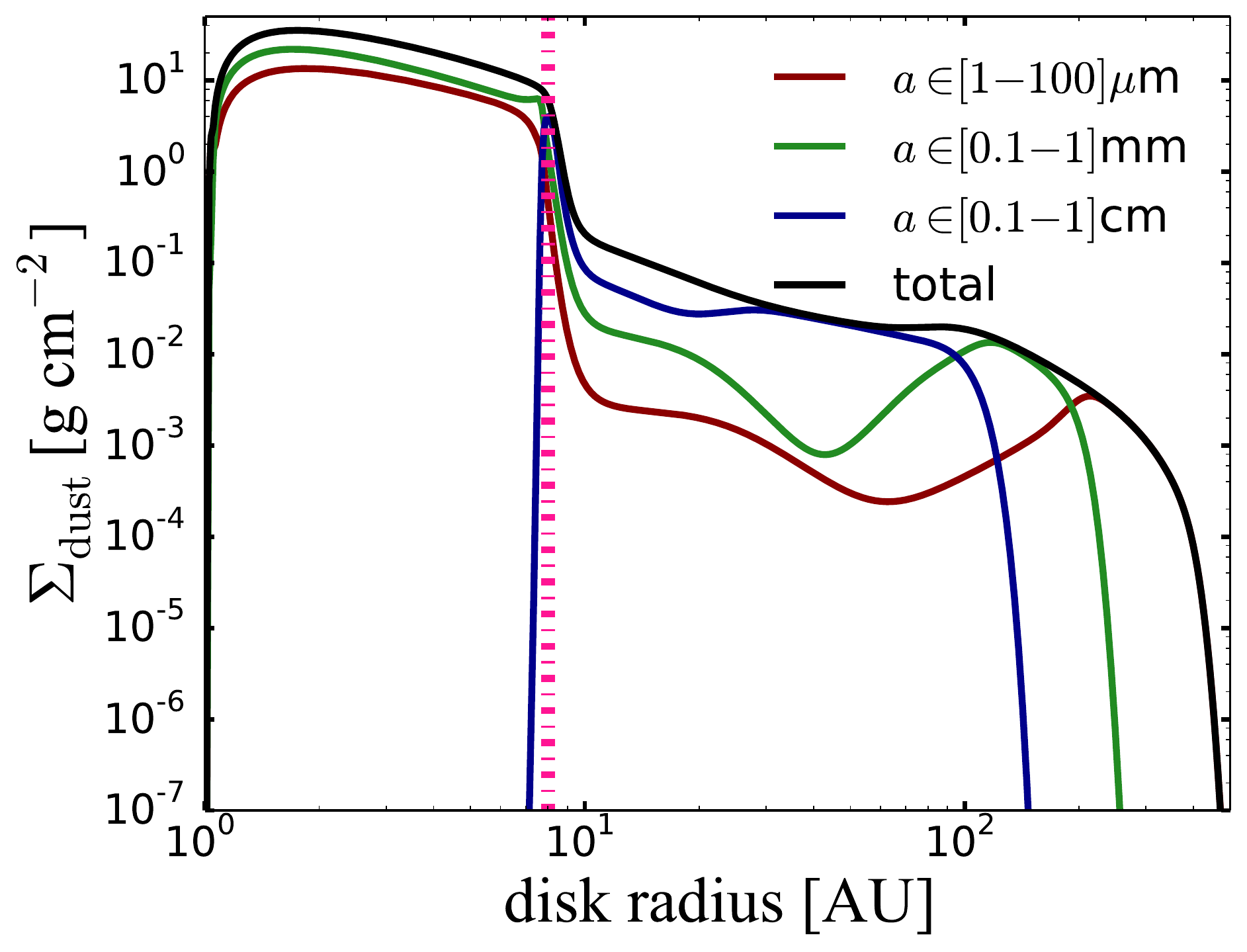}
\includegraphics[scale=0.3]{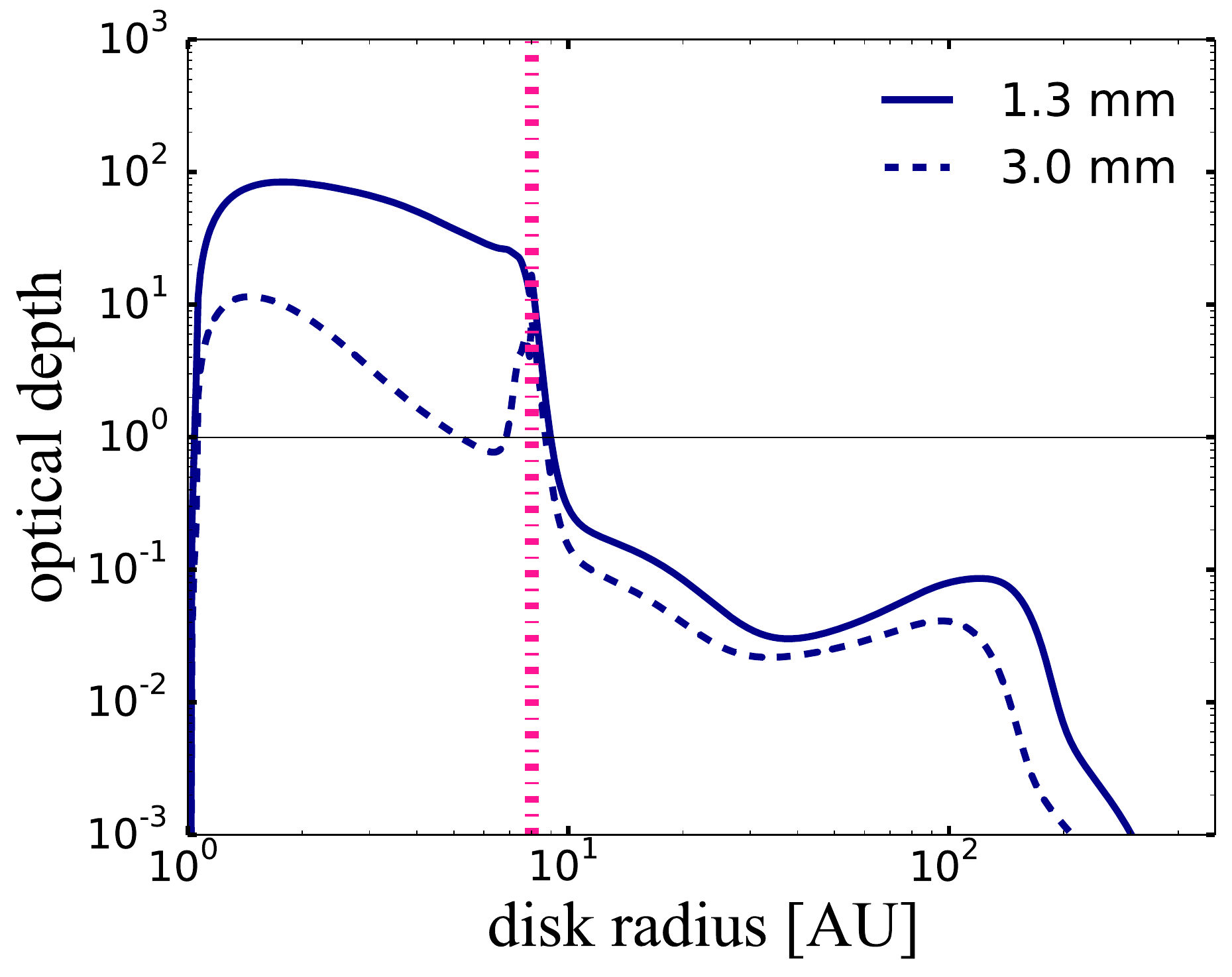}
\includegraphics[scale=0.3]{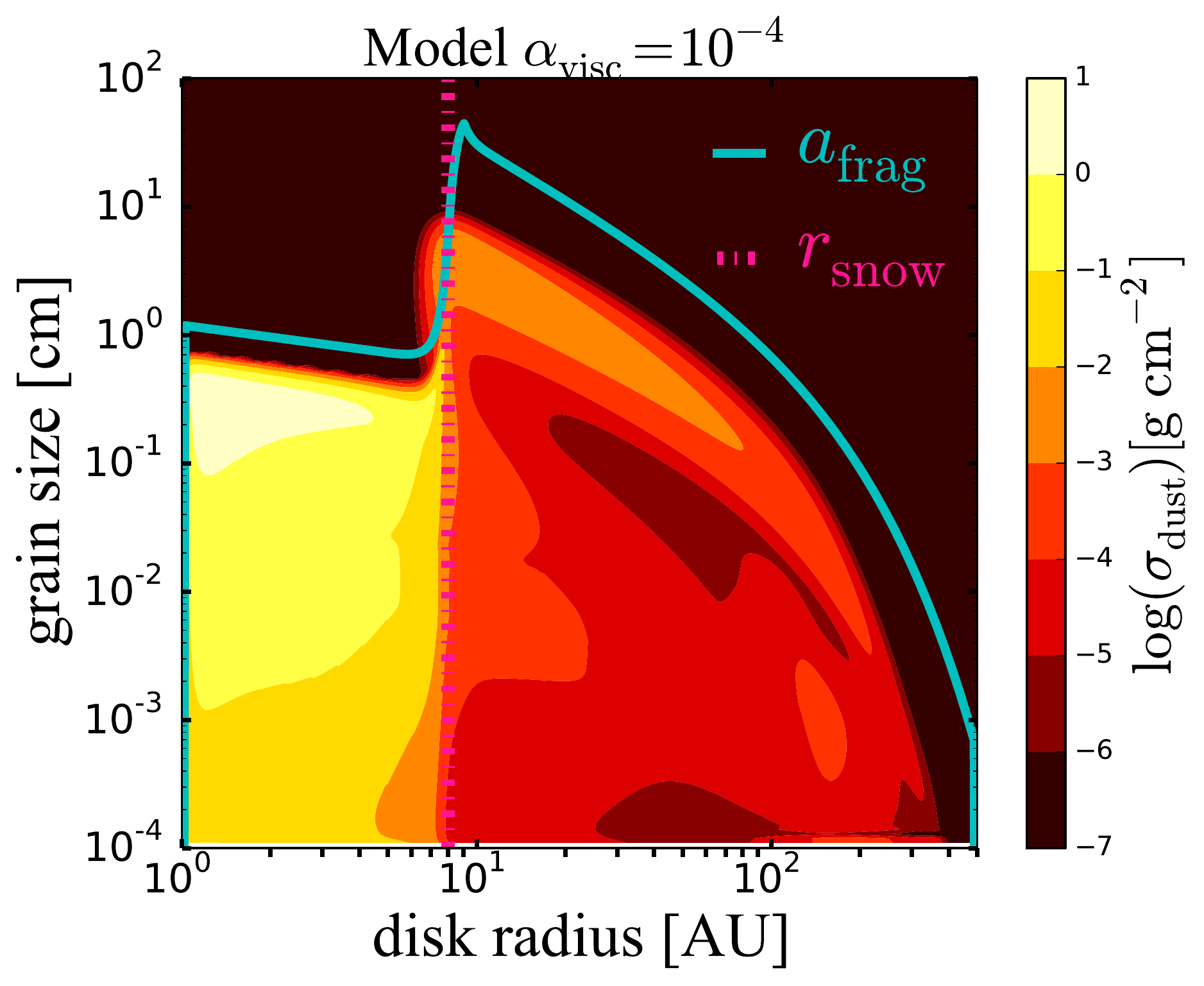}
\includegraphics[scale=0.3]{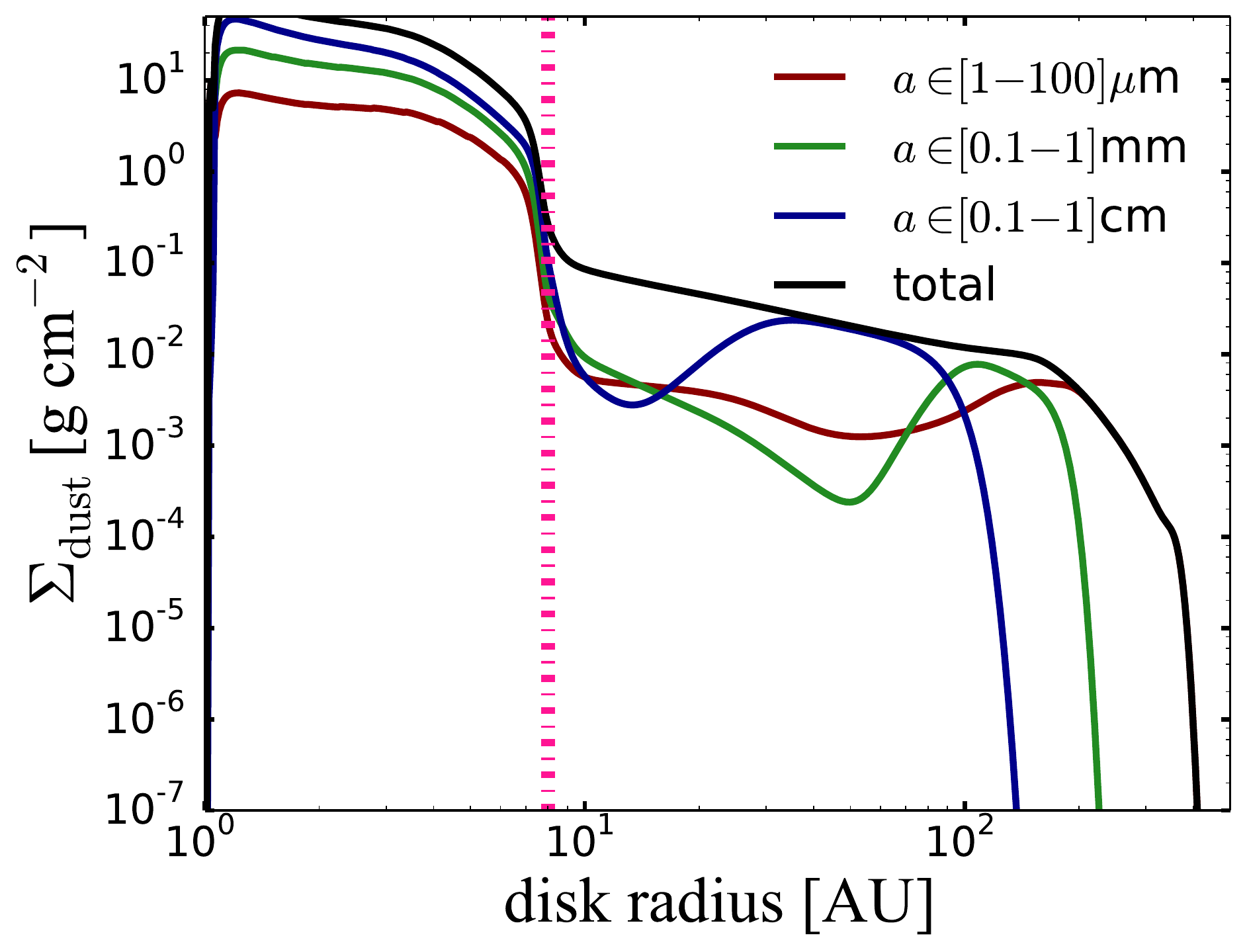}
\includegraphics[scale=0.3]{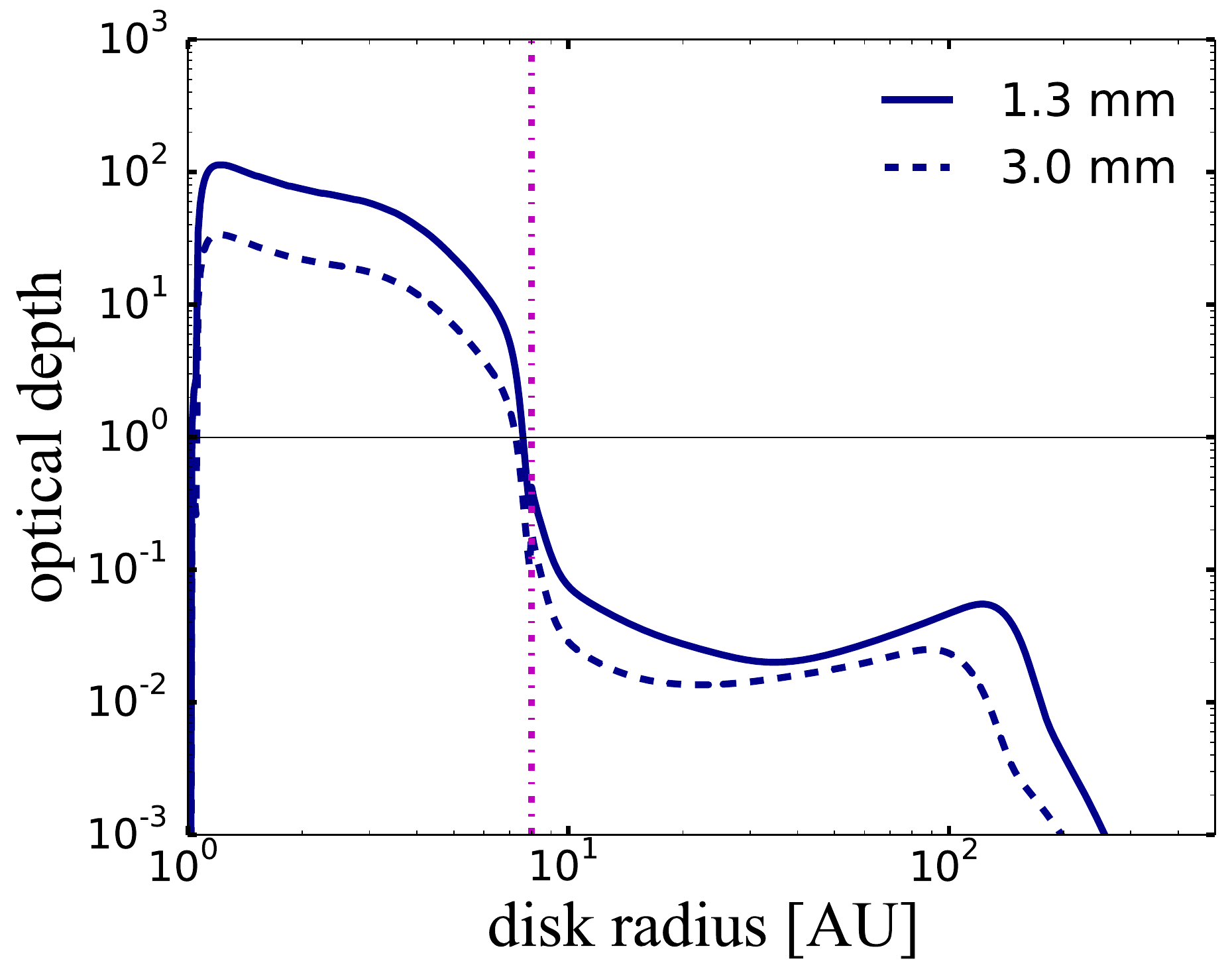}
\caption{Models with: $\alpha_{\rm{visc}} = 10^{-2}$ (\textit{top}), $\alpha_{\rm{visc}} = 10^{-3}$ (\textit{middle}), $\alpha_{\rm{visc}} = 10^{-4}$ (\textit{bottom}). For each model, we show the vertically-integrated dust density distribution (\textit{left}), the dust surface density in three grain size bins (\textit{middle}), and the optical depth of the thermal dust emission at 1.3 and 3.0 mm (\textit{right}) as a function of disk radius. The snow line radius is marked with a vertical line.}
\label{fig:grainsize}
\end{figure*}

\begin{figure*}[ht]
\includegraphics[scale=0.5]{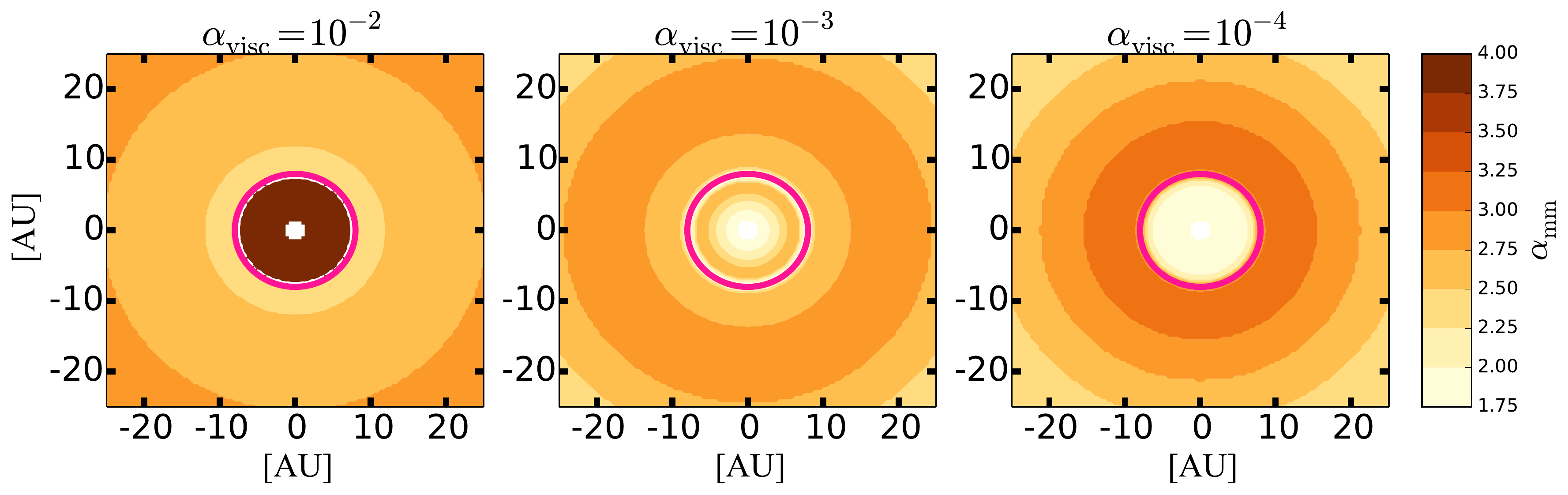}
\includegraphics[scale=0.571]{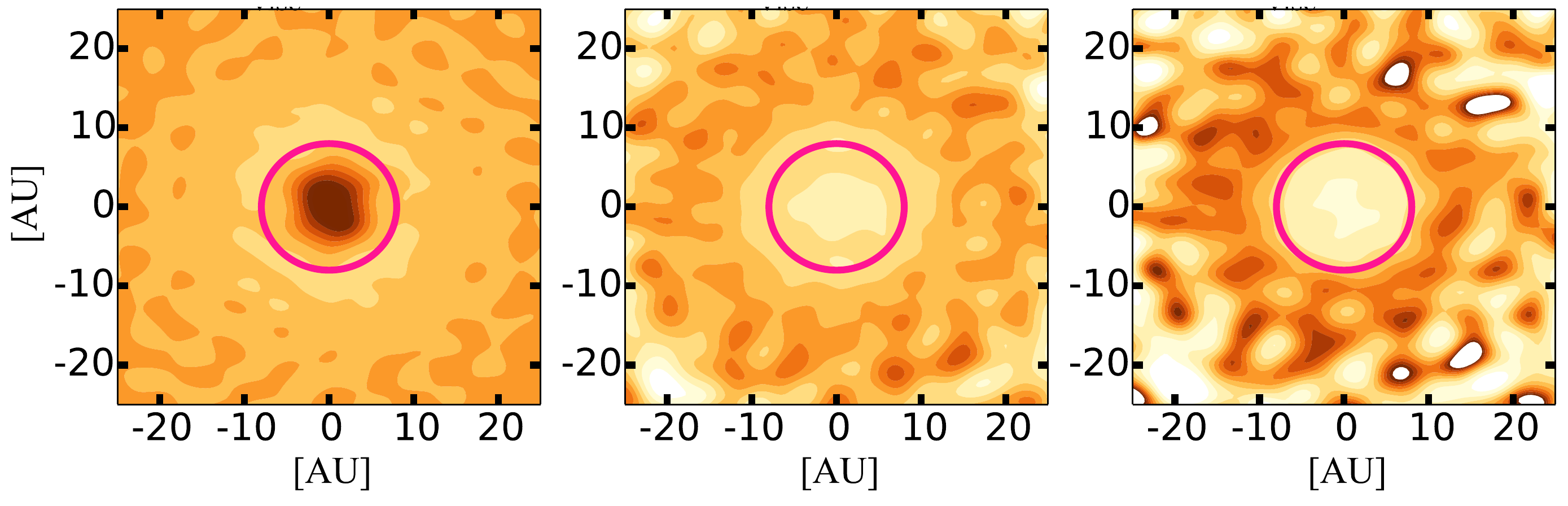}
\includegraphics[scale=0.415]{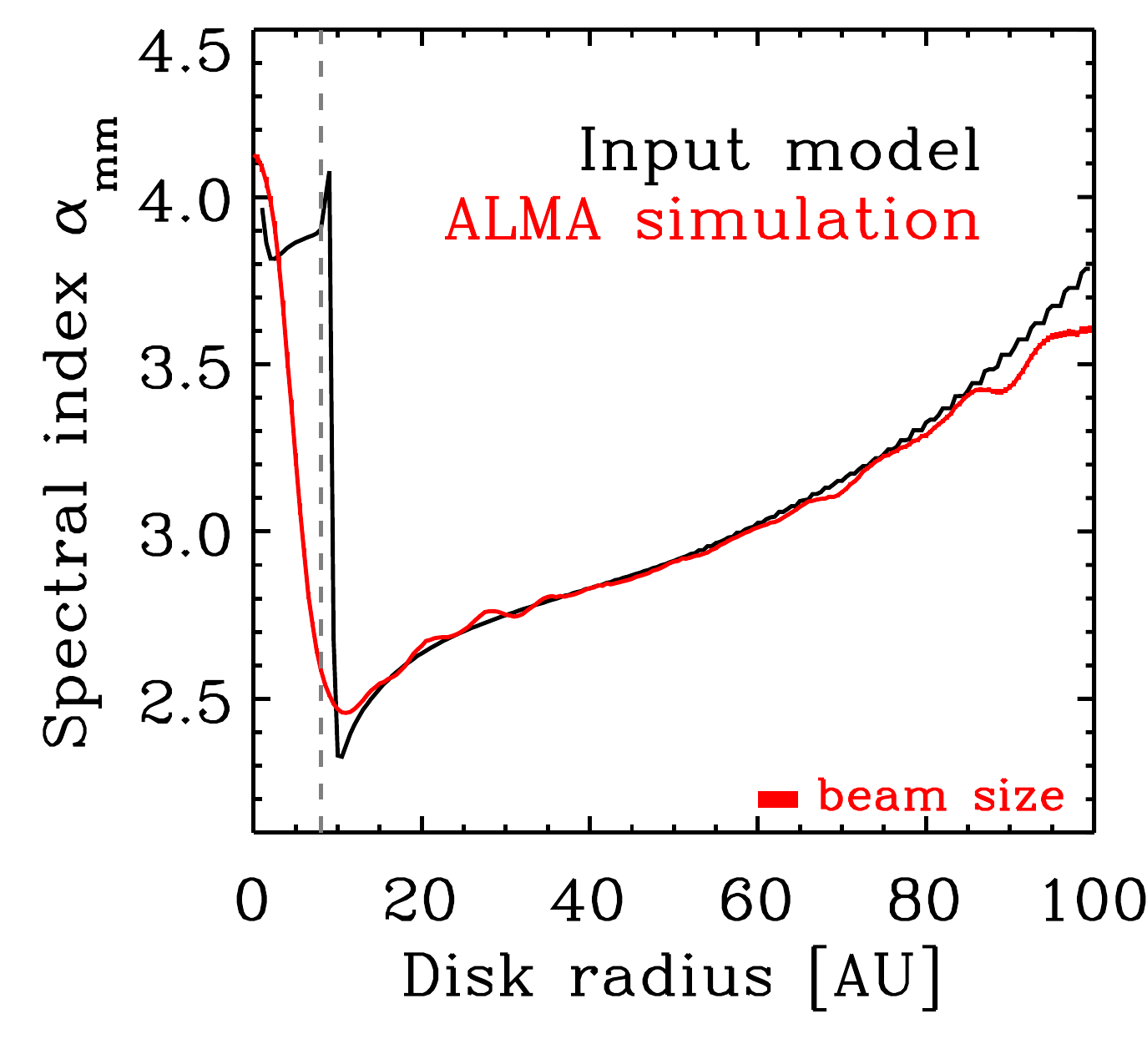}
\includegraphics[scale=0.415]{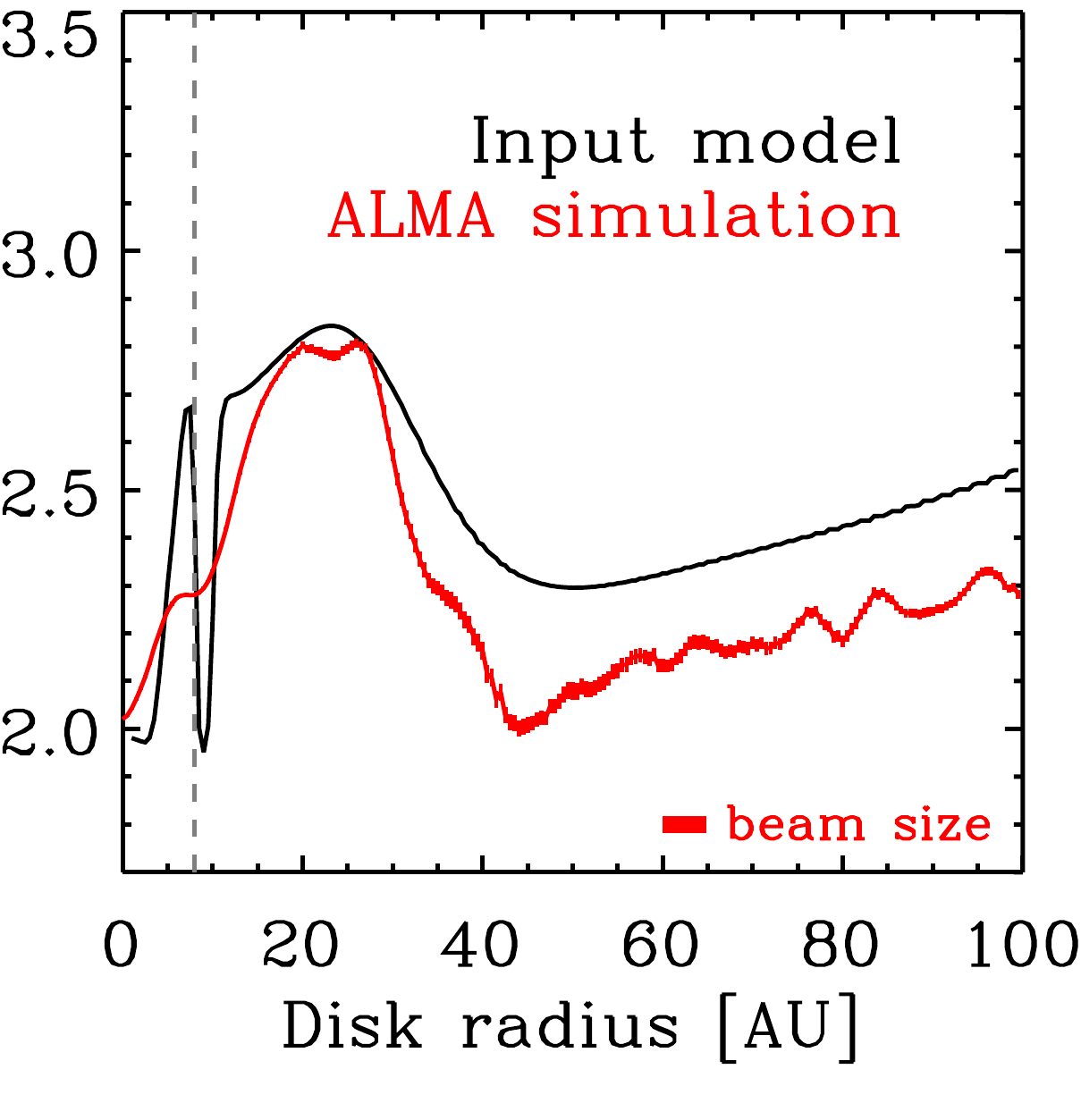}
\includegraphics[scale=0.415]{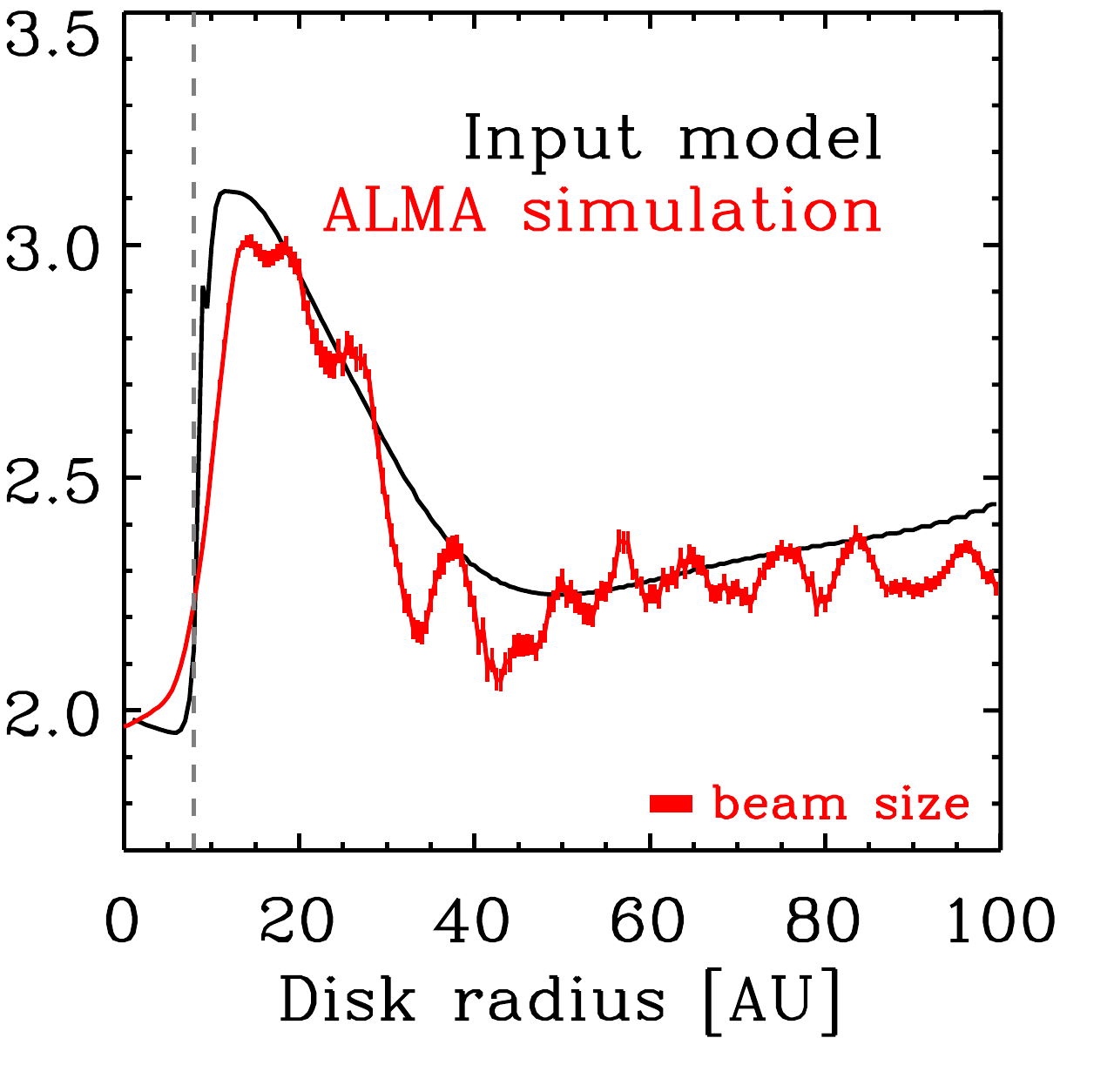}
\caption{\textit{Top}: Maps of the spectral index between 1.3 and 3.0 mm for models using different values of the alpha viscosity parameter (from left to right: $\alpha_{\rm{visc}} = 10^{-2}$, $\alpha_{\rm{visc}} = 10^{-3}$, $\alpha_{\rm{visc}} = 10^{-4}$). \textit{Middle}: ALMA-simulated spectral index maps of the models shown above. \textit{Bottom}: azimuthally-averaged radial profiles of the spectral index maps. The snow line radius is marked with a vertical line.}
\label{fig:almasim}
\end{figure*}

In the models, the gas surface density and the disk midplane temperature are taken to be constant with time and assumed to be an exponential tapered and a power law functions, respectively, as 
\begin{equation} \label{eqn: sigma_prof}
\Sigma(r) = \Sigma_c \left(\frac{r}{r_c} \right)^{-\gamma} \times \exp\left[ \left( \frac{r}{r_c}\right)^{2.0-\gamma} \right] \,,
\end{equation}
\begin{equation} \label{eqn: temp_prof}
T(r) = T_{\star}\times \left(\frac{R_{\star}}{r}\right)^{0.5}\times\phi_{\rm{inc}}^{0.25} \,.
\end{equation}
Assuming a Herbig star with $T_{\star} \sim 9000\,$K, $R_{\star} \sim 2 R_{\odot}$, and $L_{\star} \sim 30 L_{\odot}$, the water snow line is set at $\sim8\,$AU in the disk ($T_{\rm{snow}}\sim 150~$K). The cut-off radius $r_c$ is taken at 115~AU, $\phi_{\rm{inc}}=0.05$, $\gamma=0.8$, and we use a total disk mass of $M_{\rm{disk}}=0.09~M_\odot$, by taking the Herbig star of HD\,163296 as reference \citep[e.g.][]{gregmons,rosen}.
For the grain compositions, we assume a mix of silicates and carbonaceous material inside the snow line and include water ice outside the snow line \citep{pol94}, with volume fractions and optical properties from \citet{ricci10,ricci10b} and \citet{banz} and references therein.

\subsection{Modeling results} \label{sec: model_results}
The prime reason why the snow line imprints a discontinuity in the radial dust properties in a disk is a change in $v_{\rm{frag}}$. Laboratory experiments and numerical simulations have shown that the velocity for destructive collisions decrease by factors of $>$10 in the absence of ices, from 15-50 m\,s$^{-1}$ for ices outside the snow line to 1-2 m\,s$^{-1}$ for silicates inside the snow line \citep[e.g.][]{paszy,wada,gundblum}. Through Equation \ref{eqn: a_max}, the change produced in terms of grain sizes is a decrease of a factor $\approx 10^2$ inside the snow line \citep[see also][]{Til10}, causing strong effects in the observed dust emission.  

In Figure \ref{fig:grainsize} we illustrate our results. As a function of disk radius, we show the vertically integrated dust density distribution (left plot), the dust surface density integrated over three representative grain size bins (middle plot), and the optical depth of the emission at three different wavelengths (right plot). In Figure \ref{fig:almasim}, for each model, we report the ALMA-simulated image and the azimuthally-averaged radial profile of the spectral index $\alpha_{\rm{mm}}$. In the following, we describe three models that illustrate extreme cases by assuming three values of $\alpha_{\rm{visc}}=[10^{-2}, 10^{-3}, 10^{-4}]$. Variations in $v_{\rm{frag}}$ and in the ice fraction beyond the snow line produce weaker effects that for brevity we do not include here. All models are run up to $\approx1$\,Myr of evolution while keeping the same initial conditions.

In the reference model (top of Figure \ref{fig:grainsize}), we adopt $\alpha_{\rm{visc}} = 10^{-2}$ and a change in fragmentation velocity of a factor of ten across the snow line (from 10 m\,s$^{-1}$ outside, to 1 m\,s$^{-1}$ inside). Under these conditions, icy grains grow up to several cm in size in the outer disk and (weakly) drift toward the star. The relative behavior of the dust surface density in the three size-bins (middle panel of Figure \ref{fig:grainsize}) illustrates the high coagulation efficiency at disk radii of 10--100 AU, where $\mu$m-size grains are depleted by coagulation into larger grains. When icy grains cross the snow line and evaporate, the fragmentation barrier decreases and the smaller silicate grains feel a weaker drift, producing a ``traffic-jam" effect and increasing the dust surface density inside the snow line \citep[a similar result was found by][]{Til10,pin15b}. As a consequence, a grain size variation of a factor $\approx50$ is produced across the snow line, with cm-size grains outside and sub-mm-size grains inside. This, in addition to the change in composition from ices to silicates, produces a sharp increase in $\alpha_{\rm{mm}}$ from 2.5 up to 4 (Figure \ref{fig:almasim}). Because the optical depth at all mm wavelengths is $\lesssim0.1$ inside the snow line, the high value of $\alpha_{\rm{mm}}$ is sensitive to the small dust grains in the optically thin regime through $\beta$ as introduced above.
 
As the value of $\alpha_{\rm{visc}}$ is decreased (middle and bottom in Figure \ref{fig:grainsize}), the maximum grain size achievable at any given disk radius and, in turn, the drift efficiency increase. Higher drift efficiencies of mm-cm-size grains are demonstrated by the larger separation between $a_{\rm{frag}}$ and the distribution of grain sizes (left panel of Figure \ref{fig:grainsize}). Inside the snow line, dust grains up to $\sim1$\,cm exist due to the efficient coagulation and drift from the outer disk region, increasing the dust surface density by up to a factor 100 as compared to the initial profile. The ``traffic-jam" effect is thus made stronger, pushing $\alpha_{\rm{mm}}$ toward the optically-thick limit of $\sim2$. With $\alpha_{\rm{visc}} = 10^{-4}$, the dust emission is so optically thick at all mm wavelengths inside the snow line ($\tau \sim 10-100$) that the radial profile of $\alpha_{\rm{mm}}$ is opposite to the reference model. In this case, the $\alpha_{\rm{mm}}$ map shows a broad ring with maximum value of 3.1 beyond the snow line, where the emission is optically thin, reflecting the dominance of small grains due to the efficient drift-loss of cm-size grains from between 10 and 20 AU (middle panel in Figure \ref{fig:grainsize}).

\section{ALMA simulated images of a water snow line} \label{sec:alma_sim}
We simulated the model disk images at 1.3 and 3.0 mm (ALMA bands 6 and 3) using the simulator included in the CASA package for ALMA data reduction, and divided their intensities to obtain a map of the spectral index
\begin{equation}
\rm \alpha_{mm} = ln(\rm{image}_{1.3mm}/image_{3.0mm}) / ln(\lambda_{3.0}/\lambda_{1.3}) \, .
\end{equation}  
We consider wavelengths $> 1\,$mm to ensure the possibility of low optical depth inside the snow line at least in one band and in one model ($\alpha_{\rm{visc}} = 10^{-2}$); at shorter wavelengths, the disk emission is optically thick in all models explored.
It is important to obtain ALMA observations with comparable beam size at both wavelengths in order to avoid artifacts due to the difference in spatial resolutions. For the sake of demonstration only, all the simulations presented here assume an integration time of 8 hours in each band.

In Figure \ref{fig:almasim}, we show $\alpha_{\rm{mm}}$ maps for the three models, all with a beam size of $\sim0\farcs04$ after deconvolution of the Fourier image from the visibilities. The extended ALMA array configurations allow to fully resolve the region inward of the snow line ($\sim0\farcs13$, using a distance of 125 pc as for the reference disk of HD\,163296), while still recovering the disk flux over large enough scales (spatial filtering takes over beyond $\sim0\farcs8-1\farcs0$ depending on the band). By adopting different weights during synthesis imaging of the data it is possible to maximize sensitivity (using natural weights) rather than spatial resolution (using uniform weights). Here we use Briggs weights, which provide an intermediate solution and give best results in retrieving the input model profiles. 

The azimuthally-averaged spectral index radial profiles shown in Figure \ref{fig:almasim} demonstrate that the discontinuity at the water snow line can be imaged by ALMA in all cases, although convolution with the obtained spatial resolution makes all profiles shallower than they really are in the input models. As the disk emission beyond the snow line becomes weaker (for $\alpha_{\rm{visc}} < 10^{-3}$),  higher noise fluctuations are seen in the ALMA simulated images beyond $\sim30$~AU, without compromising the signal at the snow line radius.

\section{Discussion} \label{sec:disc}

In this work, we have modeled the effects of a water snow line on the properties of the dust grain population in a protoplanetary disk, accounting for coagulation, fragmentation, and transport of dust grains during evolution of the disk. We find a large variation in dust grain sizes as due to the change in fragmentation velocities across the snow line, as originally found by \citet{Til10}, where icy grains coagulate more efficiently beyond and dry silicate grains fragment more efficiently inward the snow line. This produces a sharp discontinuity in the radial profile of the dust emission spectral index $\alpha_{\rm{mm}}$ that can be observed with ALMA using the dust continuum emission measured in two bands. We find this effect persistent through the range of disk and dust grains parameters that are currently considered canonical. We propose that ALMA images of disks should be found to commonly show the water snow line, when the necessary spatial resolution is achieved.

Since the maximum grain size and therefore the drift efficiency depend on the disk viscosity, we have explored the effects of assuming different $\alpha_{\rm{visc}}$ values. Essentially, the models shown in Figures \ref{fig:grainsize} and \ref{fig:almasim} illustrate how the snow line would be imaged when the emission is optically thin or thick in both bands. From the observational point of view, these extreme cases are expected to apply broadly to the range of conditions that protoplanetary disks may provide during their evolution from thick to thin inner disks \citep[e.g.][]{alex14}. 

In terms of directly imaging the radius of the water snow line using ALMA, our simulations show a range of situations. The easiest case to interpret is the model with $\alpha_{\rm{visc}} = 10^{-2}$, where the snow line is unambiguously located by a inner disk region of high spectral index values. In this case, the optically thin emission in both bands is sensitive to the grain properties through their opacity, and $\alpha_{\rm{mm}}$ increases as driven by the higher $\beta$ from the small grains. However, our modeling shows that when the disk viscosity is lower ($\alpha_{\rm{visc}} \lesssim 10^{-3}$) the region inward of the snow line becomes more optically thick due to increasing dust surface densities driven by the efficient radial drift. In this case, the snow line would be seen as a ring in the radial profile of the spectral index, with a central region of $\alpha_{\rm{mm}} \sim 2$ where the emission becomes optically thick in both bands. Somewhat similar ring-like features may be produced by disk-carving protoplanets \citep[e.g.][]{pin15}, and may require detailed modeling of both dust and gas surface densities to distinguish between the competing scenarios. 

\begin{figure}
\centering
\includegraphics[width=0.45\textwidth]{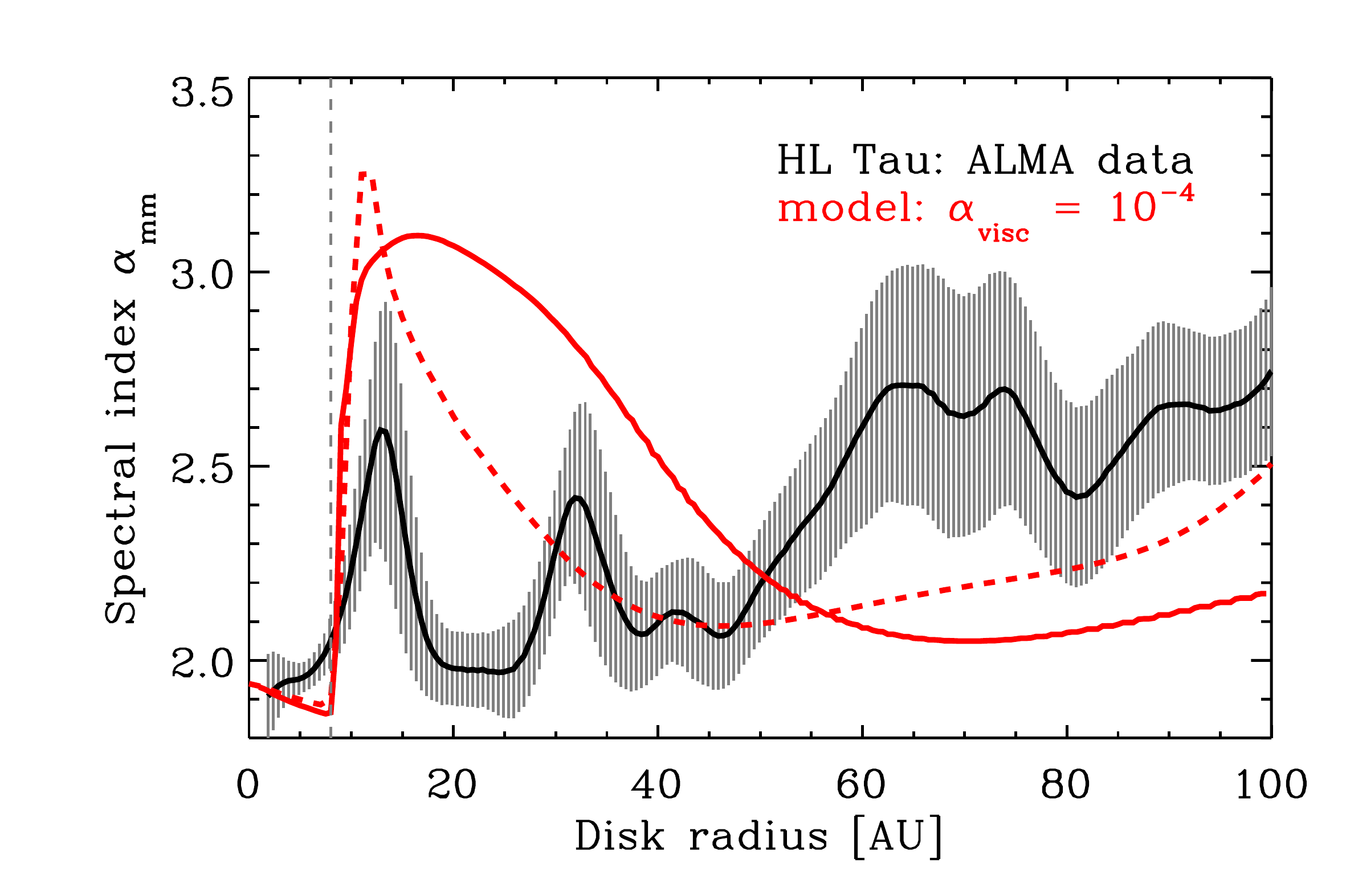} 
\caption{Our model with $\alpha_{\rm{visc}} = 10^{-4}$ overplotted to the radial profile of the spectral index between 0.87 and 1.3 mm as measured in HL Tau using ALMA \citep[grey-shaded area,][data courtesy of K. Zhang]{alma15,zhang15}. The model is shown with the temperature profile used in this work (Eqn. \ref{eqn: temp_prof}, solid line) and the higher temperature (dashed line) used in \citet{zhang15}.}
\label{fig:HLTau}
\end{figure}

The disk images of HL\,Tau taken during Science Verification of the largest ALMA array configurations provide the first (and currently the only available) example of a series of rings in the $\alpha_{\rm{mm}}$ map \citep{alma15}. A snow line interpretation of these rings has been proposed by \citet{zhang15}. They argued that each ring may correspond to the condensation front of a different molecule, with water producing the ring at $\sim6-20\,$AU. Figure \ref{fig:HLTau} compares the spectral index profile from our $\alpha_{\rm{visc}} = 10^{-4}$ model to the profile measured in HL\,Tau. The model does produce a qualitatively similar trend of $\alpha_{\rm{mm}} = 2$ inside the snow line and higher outside, but the model ring is broader than those in HL\,Tau. A narrower ring is produced in our model by using a higher midplane temperature profile, as used in \citet{zhang15}\footnote{They assume a midplane temperature profile of $T = 665(r/\rm{AU})^{-0.6}$~K from \citet{men99}.}. This happens because a higher temperature increases the dust drift velocities and lowers the maximum grain sizes, as $c_{s} \propto \sqrt{T}$ in Eqn.\ref{eqn: a_max} \citep{Til10}. Within a snow line interpretation, our model would suggest a disk viscosity as low as $\alpha_{\rm{visc}} \lesssim 10^{-3}$; this would be consistent with results from the detailed radiative transfer modeling of HL\,Tau by \citet{pinte15}. However, simple estimates by \citet{okoz15} using the ALMA data provide a temperature profile that is $\sim2$ times lower than that used by \citet{zhang15}, and the water snow line in HL\,Tau may in fact yet be undiscovered in the unresolved region at $\sim3.5$~AU. It is yet also unclear whether a significant variation in fragmentation velocity may be produced by other ices, to test within our model the interpretation that all $\alpha_{\rm{mm}}$ rings observed in HL\,Tau may be linked to snow lines \citep{zhang15,okoz15} rather than planets \citep[e.g.][]{dong15}. With an ALMA continuum survey of snow lines in disks it will be possible to test these ideas in the context of scaling relations with stellar masses and luminosities.
\\

We thank the referee for helpful feedback. Andrea Banzatti acknowledges financial support by a NASA Origins of the Solar System Grant No. OSS 11-OSS11-0120, a NASA Planetary Geology and Geophysics Program under grant NAG 5-10201. Paola Pinilla is supported by Koninklijke Nederlandse Akademie van Wetenschappen professor prize to Ewine van Dishoeck.

\end{document}